\documentclass{PoS}

\title{The Constrained E$_6$SSM}

\ShortTitle{The Constrained E$_6$SSM}

\author{Peter ATHRON\\
        Institut f\"ur Kern- und Teilchenphysik, TU Dresden, D-01062, Germany\\
        E-mail: \email{p.athron@physik.tu-dresden.de}}

\author{Stephen F.\ KING\\
        School of Physics and Astronomy, University of Southampton, Southampton, SO17 1BJ, U.K.\\
        E-mail: \email{sfk@hep.phys.soton.ac.uk}}

\author{\speaker{David J.\ MILLER}\\
        Department of Physics and Astronomy, University of Glasgow, Glasgow G12 8QQ, U.K.\\
        E-mail: \email{d.miller@physics.gla.ac.uk}}

\author{Stefano MORETTI\\
        School of Physics and Astronomy, University of Southampton, Southampton, SO17 1BJ, U.K.\\
	{\it and}\\
	Dipartimento di Fisica Teorica, Universit\`a di Torino, Via Pietro Giuria 1, 10125 Torino, Italy.\\
        E-mail: \email{stefano@phys.soton.ac.uk}}

\author{Roman NEVZOROV
        \thanks{On leave of absence from the Theory Department,
ITEP, Moscow, Russia}\\
        Department of Physics and Astronomy, University of Glasgow, Glasgow G12 8QQ, U.K.\\
        E-mail: \email{r.nevzorov@physics.gla.ac.uk}}

\abstract{

We discuss the predictions of a constrained version of the exceptional
supersymmetric standard model (cE6SSM), with a universal high energy
soft scalar mass, soft trilinear coupling and soft gaugino mass. The
spectrum includes a light gluino, a light wino-like neutralino and
chargino pair and a light bino-like neutralino, with other sparticle
masses except the lighter stop being much heavier. We also discuss
scenarios with an extra light exotic colour triplet of fermions and
scalars and a TeV scale $Z^\prime$, which lead to early exotic physics
signals at the LHC.}

\FullConference{European Physical Society Europhysics Conference on High Energy Physics\\
  July 16-22, 2009\\
  Krakow, Poland}

\begin{document}

The Exceptional Supersymmetric Standard Model (E$_6$SSM)
\cite{King:2005jy} is inspired by Grand Unification under the gauge
group $E_6$. It does not contain the $E_6$ symmetry, but supposes a
breaking $E_6 \to SO(10)\times U(1)_{\psi}$, followed by $SO(10)\to
SU(5)\times U(1)_{\chi}$, to give the subgroup $SU(3)_C\times
SU(2)_L\times U(1)_Y\times U(1)'$ at low energies. To allow heavy
right-handed neutrinos (facilitating the see-saw mechanism), the
remaining $U(1)'$ at low energy is the combination $U(1)'=U(1)_{\chi}
\cos\theta+U(1)_{\psi} \sin\theta$ with $\theta=\arctan\sqrt{15}$,
which keeps the right-handed neutrinos sterile.

The particle content forms a complete $27$
representation of $E_6$ for each generation, cancelling
anomalies automatically. The $27_i$ decomposes under the
$SU(5)\times U(1)_{N}$ subgroup of $E_6$ as
\begin{equation}
27_i \to \left(10,\,1\right)_i+\left(5^{*},\,2\right)_i
+\left(5^{*},\,-3\right)_i +\left(5,-2\right)_i
+\left(1,5\right)_i+\left(1,0\right)_i\,,
\label{4}
\end{equation}
where bracketed quantities are the $SU(5)$ representation
and $U(1)_{N}$ charge ($\times \sqrt{40}$) with $i$ a family
index. Ordinary SM quarks and leptons are assigned to
$\left(10,1\right)_i+\left(5^{*},\,2\right)_i$, right-handed neutrinos
$N^c_i$ appear in $\left(1,0\right)_i$, and $\left(1,5\right)_i$
provides singlet fields ($S_i$) that carry non-zero $U(1)_{N}$ charges
and survive down to the EW scale.  There are three pairs of
$SU(2)$-doublets ($H^d_{i}$ and $H^u_{i}$) in
$\left(5^{*},\,-3\right)_i$ and we identify the third generation with
the MSSM Higgs doublets; the other two do not get VEVs and we refer to
them as ``inert''. These multiplets also contain colour triplets of
exotic quarks ($D_i$ and $\overline{D_i}$).  The model also requires
superfields $H'$ and $\overline{H}'$ from extra incomplete
representations to ensure gauge coupling unification (see
Ref.~\cite{Howl:2008xz} for an alternate unification scenario).

Extra symmetries are required to prevent unwanted flavour changing
neutral currents and proton decay.  To suppress flavour changing
neutral currents, one postulates a $Z^{H}_2$ symmetry under which all
superfields except the third generation Higgs fields are odd. This can
only be an approximate symmetry, otherwise the exotics would not be
able to decay.  To prevent rapid proton decay, a generalisation of
R-parity is imposed. If the Higgs, exotic quarks and quark superfields
are even under a discrete $Z^L_2$ symmetry while the lepton
superfields are odd, we will call this Model I and the superpotential
is invariant with respect to a $U(1)_B$ global symmetry. The exotic
$\overline{D_i}$ and $D_i$ are then identified as diquark and
anti-diquark, i.e. $B_{D}=-2/3$ and
$B_{\overline{D}}=2/3$. Alternatively (model II), the exotic quarks
and the lepton superfields could be odd under $Z^B_2$ whereas the
others superfields remain even. In this case the $\overline{D_i}$ and
$D_i$ are leptoquarks.

Integrating out the heavy right-handed neutrinos, and assuming a hierarchical structure of Yukawas, the superpotential becomes approximately, 
\begin{equation}
W_{\rm E_6SSM} \simeq \lambda_i
S_3(H^d_i H^u_i)+ \kappa_i S_3 (D_i\overline{D}_i) 
+h_t(H_3^{u}Q)t^c+h_b(H_3^{d}Q)b^c+ h_{\tau}(H_3^{d}L)\tau^c+
\mu'(H^{'}\overline{H^{'}}),
\label{cessm8}
\end{equation}
where $i=1,2,3$ and $\lambda_i$, $\kappa_i$ are dimensionless
couplings. The above
ignores small $Z_2^H$ violating terms such as
$g_{ijk} D_i \left( Q_j Q_k \right)$ which are unimportant for
production, but may be important for decay.\\[-2mm]

The E$_6$SSM has 43 new parameters (14 of which are phases) compared
to the MSSM, so it is useful to consider a constrained model, where
the parameters unify at the GUT scale, $M_X$. In the constrained
E$_6$SSM (cE$_6$SSM), soft scalar masses are set to $m_0$, gaugino
masses are set to $M_{1/2}$ and trilinear scalar couplings are set to
$A_0$, at the scale $M_X$~\cite{Athron:2009ue,Athron:2009bs}.
It is thus characterised by the parameters
$\lambda_i(M_X),\, \kappa_i(M_X),\, h_t(M_X),\, h_b(M_X), \, h_{\tau}(M_X), \, m_0, \, M_{1/2}$ and $A_0$.

To calculate the low energy spectrum, we derived two-loop
renormalisation group equations (RGEs) for gauge and Yukawa couplings,
two--loop RGEs for $M_a(Q)$ and $A_i(Q)$ and one--loop RGEs for
$m_i^2(Q)$. We implemented these into a modified version of SOFTSUSY
2.0.5 \cite{Allanach:2001kg}. The gauge and Yukawa couplings are
independent of the soft SUSY breaking parameters so are determined
first. The Higgs VEVs, third generation Yukawas and gauge couplings
(except $g_i^\prime$) are input at the low energy scale, while
$\kappa_i(M_X)$, $\lambda_i(M_X)$ are high scale inputs. We insist on
gauge coupling unification, and iterate until everything is
consistent.  The low energy SUSY breaking parameters are determined
semi-analytically as (linear or quadratic) functions of $A_0$,
$M_{1/2}$ and $m_0$. The coefficients of each term are unknown
analytically but are determined numerically. We obtain values for
$m_0$, $M_{1/2}$ and $A_0$ that are consistent with EWSB by imposing
minimisation conditions on the one-loop effective Higgs potential
(obtaining up to four solutions for each set of Yukawa couplings), and
then calculate the mass spectrum.  Although correct EWSB is not
guaranteed, there are always solutions with real $A_0$, $M_{1/2}$ and
$m_0$ for sufficiently large $\kappa_i$. $\kappa_i$ couples the
singlet to a large number of coloured fields, efficiently driving its
squared mass negative to trigger symmetry breaking.

To avoid conflict with experiment we impose: $m_h \geq 114$ GeV;
sleptons and charginos are heavier than $100\,\mbox{GeV}$; squarks and
gluinos have masses above $300\,\mbox{GeV}$ and the $Z'$ boson has a
mass larger than $861\,\mbox{GeV}$. We also impose bounds on the
exotic quarks and squarks from the HERA experiments, by requiring them
to be heavier than $300\,\mbox{GeV}$. We require the inert Higgs and
inert Higgsinos are heavier than 100 GeV due to LEP bounds.  Finally,
we impose some theoretical constraints: the lightest SUSY particle
(LSP) should be a neutralino and we restrict the GUT scale Yukawa
couplings to be less than 3 to ensure applicability of perturbation
theory. \\[-2mm]

To investigate the phenomenology, we fixed values of $\tan \beta =
3,\,10,\,30$ and scanned over $s,\,\kappa,\,\lambda$, determining
values of the soft mass parameters $A_0,\,M_{1/2}$ and $m_0$
consistent with the correct breakdown of electroweak symmetry. We
found that $m_0>M_{1/2}$ for each value of $s$ and also that lower
$M_{1/2}$ is weakly correlated with lower $s$ and thus lower $Z'$
masses. The low energy gaugino masses are driven by RG running to be
small, so the lightest SUSY states are generally a light gluino of
mass $\sim M_3$, a light wino-like neutralino and chargino pair of
mass $\sim M_2$, and a light bino-like neutralino of mass $\sim M_1$,
which are typically all much lighter than the Higgsino masses of order
$\mu = \lambda \langle S_3 \rangle$.  The squarks and sleptons are
much heavier than these light gauginos.

\begin{figure}[h]
 \begin{center}
\vspace{-3mm}
\resizebox{!}{5.8cm}
{\includegraphics{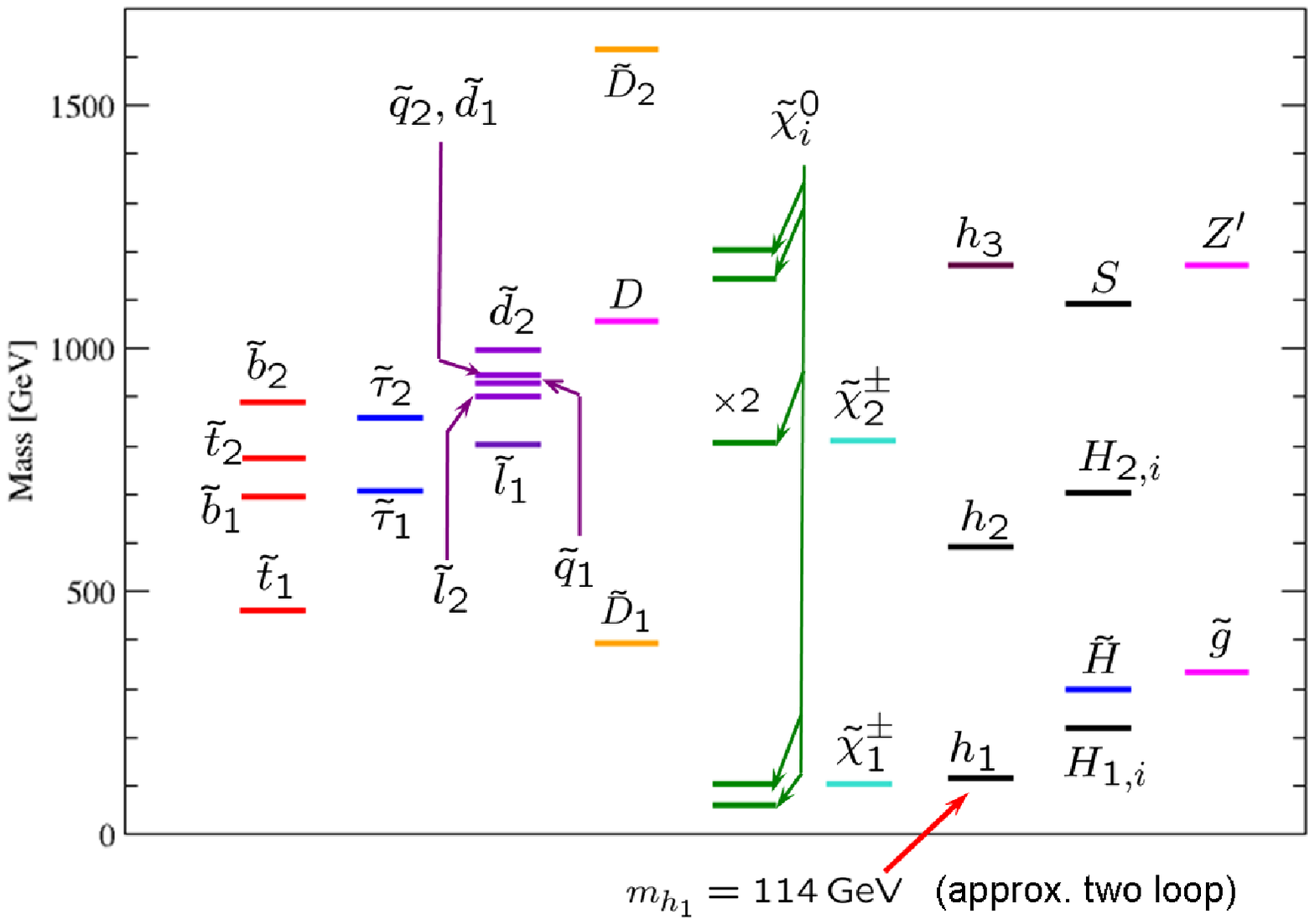}}
\resizebox{!}{5.8cm}
{\includegraphics{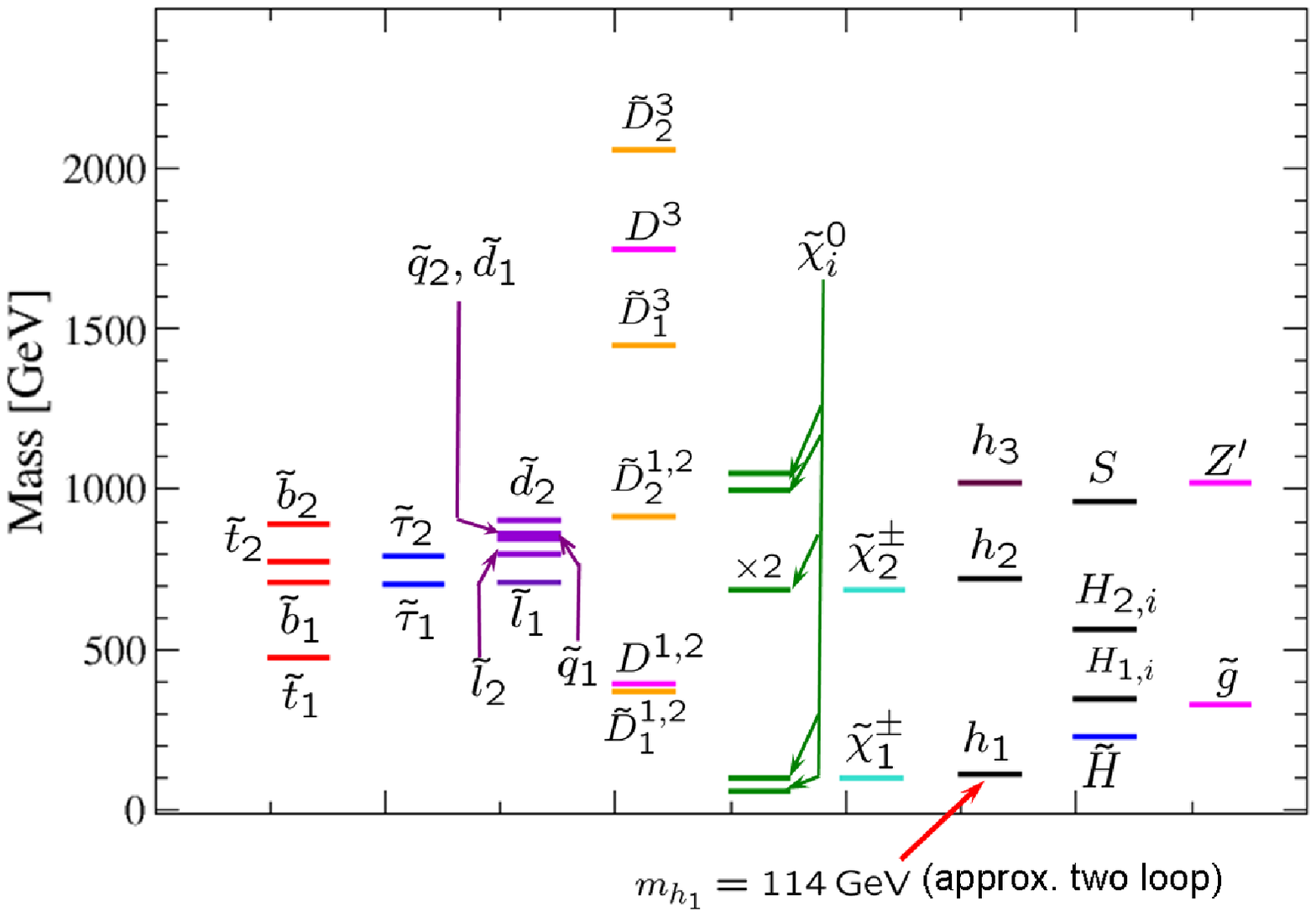}}
\vspace{-3mm}
\caption{Spectra for two cE$_6$SSM scenarios. For explicit parameters, 
see Ref.\cite{Athron:2009ue}
\label{benchmarks}}
\vspace{-3mm}
\end{center}
\end{figure}

We show two possible particle spectra in Fig.~\ref{benchmarks} (see
Ref.~\cite{Athron:2009ue} for the explicit parameters used).
The light gauginos ensure that pair production of $\chi_2^0\chi_2^0$,
$\chi_2^0\chi_1^\pm$, $\chi_1^\pm \chi_1^\mp$ and $\tilde g \tilde g$
should always be possible at the LHC.  The gluinos are relatively
narrow states with width $\propto M_{\tilde{g}}^5/m_{\tilde{q}}^4$,
and decay via $\tilde{g} \rightarrow q \tilde{q}^*
\rightarrow q \bar{q} + E_T^{\rm miss}$, resulting in an
enhancement of the cross section $pp \rightarrow q \bar q q \bar q
+ E_T^{\rm miss} + X$.

For the left-hand scenario, we have taken $\kappa_1=\kappa_2=\kappa_3$
at the GUT scale and large enough to trigger EWSB, resulting in heavy
and degenerate exotic quarks. Nevertheless, due to mixing, we find a
light exotic coloured scalar with mass $393 \, {\rm GeV}$.  For the
right-hand scenario, we have instead set $\kappa_3 \gg \kappa_{1,2}$,
allowing for rather light exotic quarks of $391 \, {\rm GeV}$.

Assuming $D_i$ and $\overline{D}_i$ couple most strongly to the third
generation, the lightest $D_i$ and $\overline{D}_i$ decay into
$\tilde{t}b$, $t\tilde{b}$, $\bar{\tilde{t}}\bar{b}$,
$\bar{t}\bar{\tilde{b}}$ (if diquarks) or $\tilde{t}\tau$,
$t\tilde{\tau}$, $\tilde{b} \nu_{\tau}$, $b\tilde{\nu_{\tau}}$ (if
leptoquarks). This leads to a substantial enhancement of either $pp\to
t\bar{t}b\bar{b}+E^{\rm miss}_{T}+X$ (if diquarks) or $pp\to
t\bar{t}\tau \bar{\tau}+E^{\rm miss}_{T}+X$ or $pp\to b\bar{b}+ E^{\rm
miss}_{T}+X$ (if leptoquarks). Therefore light leptoquarks should
produce a strong signal with low SM background at the LHC. For
example, for a $D$ mass of $400 \, {\rm GeV}$, the production
cross-section $pp \to D \bar D$ is $17.4 \, {\rm pb}$. The observation
of the D fermions should be possible if they have masses below about
1.5-2 TeV \cite{King:2005jy}.  The exotic scalars ($\tilde{D}_i$ and
$\tilde{\overline{D}}_i$) may be produced singly and decay into
quark--quark (if diquarks) or quark--lepton (if leptoquarks) without
missing energy.  Recent Tevatron searches for dijet resonances rule
out scalar diquarks with mass less than $630 \, {\rm GeV}$. However,
scalar leptoquarks may be as light as $300 \, {\rm GeV}$ since they
are pair produced through gluon fusion.  They then decay, e.g.\
$\tilde D \rightarrow t \tau$, leading to an enhancement of $pp
\rightarrow t \bar t \tau \bar{\tau}$.

A light inert Higgs decays via terms analogous to the Yukawa
interactions of normal Higgs superfields, leading to decays such as
$H^{0}_{1,\,i} \rightarrow b \bar b$ and $H^{-}_{1,\,i} \rightarrow
\tau \bar{\nu}_{\tau}$.  Similar couplings govern inert Higgsino
decays resulting in e.g.\ $\tilde H_{i}^0 \rightarrow t
\tilde{\bar{t}}^*$, $\tilde H_{i}^0 \rightarrow \tau
\tilde{\bar{\tau}}^*$, $\tilde H_{i}^{+} \rightarrow t
\tilde{\bar{b}}^*$ and $\tilde H_{i}^{-} \rightarrow \tau
\tilde{\bar{\nu}}_{\tau}^*$.

To conclude, the cE$_6$SSM is a well motivated model with a
distinctive spectrum: a light gluino (much lighter than the
squarks), and new exotic states such as a $Z'$ and colour triplet
fermions and scalars. If these new states are light enough, they could
be found with early LHC data, leading to a revolution in particle
physics, and pointing towards an underlying high energy $E_6$ gauge
structure.\\[-2mm]

\noindent {\bf Acknowledgements:} DJM acknowledges support from the
STFC Advanced Fellowship Grant PP/C502722/1. RN acknowledges support
from the SHEFC grant HR03020 SUPA 36878. SFK acknowledges partial
support from: STFC Rolling Grant ST/G000557/1 (also SM); EU Network
MRTN-CT-2004-503369; NATO grant PST.CLG.980066 (also SM); EU ILIAS
RII3-CT-2004-506222. SM is also supported in part by the FP7 RTN
MRTN-CT-2006-035505, and the scheme ``Visiting Professor - Azione D -
Atto Integrativo tra la Regione Piemonte e gli Atenei Piemontesi''.

\end{document}